\documentclass[fleqn,usenatbib]{mnras}

\usepackage{newtxtext,newtxmath}

\usepackage[T1]{fontenc}
\usepackage{ae,aecompl}

\usepackage{graphicx}	
\usepackage{amsmath}	
\usepackage{amssymb}	

\newcommand{\photu}{photon units}

\newcommand{\galex}{{\it GALEX}}

\newcommand{\voyager}{{\it Voyager}}

\title{The Diffuse Ultraviolet Background Close to the Galactic Plane}

\author[Jayant Murthy et al.]{
Jayant Murthy,$^{1}$\thanks{E-mail: jmurthy@yahoo.in}
R. C. Henry,$^{2}$\thanks{E-mail: henry@jhu.edu}
James Overduin$^{3}$\thanks{E-mail: joverduin@towson.edu}
\\
$^{1}$Indian Institute of Astrophysics, Bengaluru 560 034, India\\
$^{2}$Henry A. Rowland Department of Physics and Astronomy, The Johns Hopkins University, Baltimore, MD 21218, USA\\
$^{3}$Department of Physics, Astronomy and Geosciences, Towson University, Towson, MD 21252, USA\\
}

\date{Accepted XXX. Received YYY; in original form ZZZ}

\pubyear{2019}

\begin{document}
\label{firstpage}
\pagerange{\pageref{firstpage}--\pageref{lastpage}}
\maketitle

\begin{abstract}
We have used \voyager\ and \galex\ observations to map the diffuse Galactic light near the Galactic equator. We find that most of the observations are relatively faint with surface brightnesses of less than 5,000 \photu. This is important because many ultraviolet telescopes have not observed at low Galactic latitudes because of the fear of a bright diffuse emission. Our data are consistent with emission from interstellar dust grains with albedo ($a$) of 0.2 -- 0.3 and phase function ($g$) $ < 0.7$ at 1100 \AA; $0.2 < a < 0.5; g < 0.8$ at 1500 \AA; and $0.4 < a < 0.6; g < 0.4$ at 2300 \AA.
\end{abstract}




\section{Introduction}

The diffuse Galactic light (DGL) in the ultraviolet (UV) is comprised of many different components whose relative contributions vary across the sky. Stellar radiation scattered from interstellar dust dominates the DGL at low Galactic latitudes \citep{Murthy_dustmodel2016} with extragalactic radiation being more important near the poles where there is little dust \citep{Akshaya2018, Akshaya2019}. Molecular hydrogen fluorescence contributes even at high latitudes \citep{Akshaya2019} and may contribute more at low latitudes where there are more molecular clouds. Unfortunately most missions capable of observing the diffuse radiation in the UV avoided the Galactic plane because of the fear of damaging the detectors due to the intense radiation expected \citep{Martin2005} and there are few observations of the DGL at low latitudes.

One of the first observations in the Galactic plane was made using the \voyager\ ultraviolet spectrometer (UVS) in the vicinity of the Coalsack Nebula where \cite{Murthy_CS1994} found intense radiation at 1100 \AA, which they attributed to the scattering of the light from 3 of the 5 brightest UV stars in the sky from a thin layer of foreground dust. Further studies with both \voyager\ and the Far Ultraviolet Spectroscopic Explorer ({\it FUSE}) by \citet{Murthy_sahnow2004} showed a patchy distribution at low latitudes with dark regions even in the Galactic Plane. The DGL in the UV is due to the light from stars scattered by nearby dust within a few hundred parsecs of the Sun and is therefore concentrated near bright stars.

We will look at the DGL using both \voyager\ and \galex\ data. Because these are archival observations made with long inoperative instruments, we have incomplete coverage and have a limited ability to confirm uncertain observations. Nevertheless, they provide valuable input on the sources of the diffuse light in the Galactic Plane.

\section{Data}
\subsection{\voyager}

\begin{table}
	\centering
	\caption{Exclusion dates for Voyager planetary encounters.}
	\label{tab:voy_exclusion}
	\begin{tabular}{ll}
		\hline
	\multicolumn{2}{c}{\voyager\ 1}\\
	\hline
		1979 -- 1979.29 & Jupiter\\
		1980.64 - 1981 & Saturn\\
		\hline
	\multicolumn{2}{c}{\voyager\ 2}\\
	\hline
		1979.3 -- 1979.9 & Jupiter\\
		1981.4 -- 1981.8 & Saturn\\
		1985.8 -- 1986.2 & Uranus\\
		1989.4 -- 1989.8 & Neptune\\
		\hline
	\end{tabular}
\end{table}

The two \voyager\ spacecraft were launched in 1977 with the primary mission objective of visiting the Jovian planets. However, during their cruise phase and after the last of their planetary encounters, the two spacecraft observed many astrophysical targets with their ultraviolet spectrometers (UVS). The two UVS are identical Wadsworth-mounted objective grating spectrometers which cover the spectral range between 500 and 1700 \AA\ (with maximum sensitivity for $\lambda < 1200 $\AA) with a field of view of $0^{\circ}.1 \times\ 0^{\circ}.87$ \citep{Broadfoot1977}. \citet{Murthy_voy, Murthyvoy_all} reprocessed the entire \voyager\ database of several thousand observations and tabulated the diffuse observations, from which we have extracted those within $10^{\circ}$ of the Galactic equator. Note that we have excluded several of the brightest points from \citet{Murthyvoy_all} which, upon further examination, were found to be near planetary encounters (Table \ref{tab:voy_exclusion}). They were flagged as observations of the blank sky but cluster around the planetary encounters and are probably observations of the planetary environment

\begin{figure*}
	\includegraphics[width=\textwidth, height=4in]{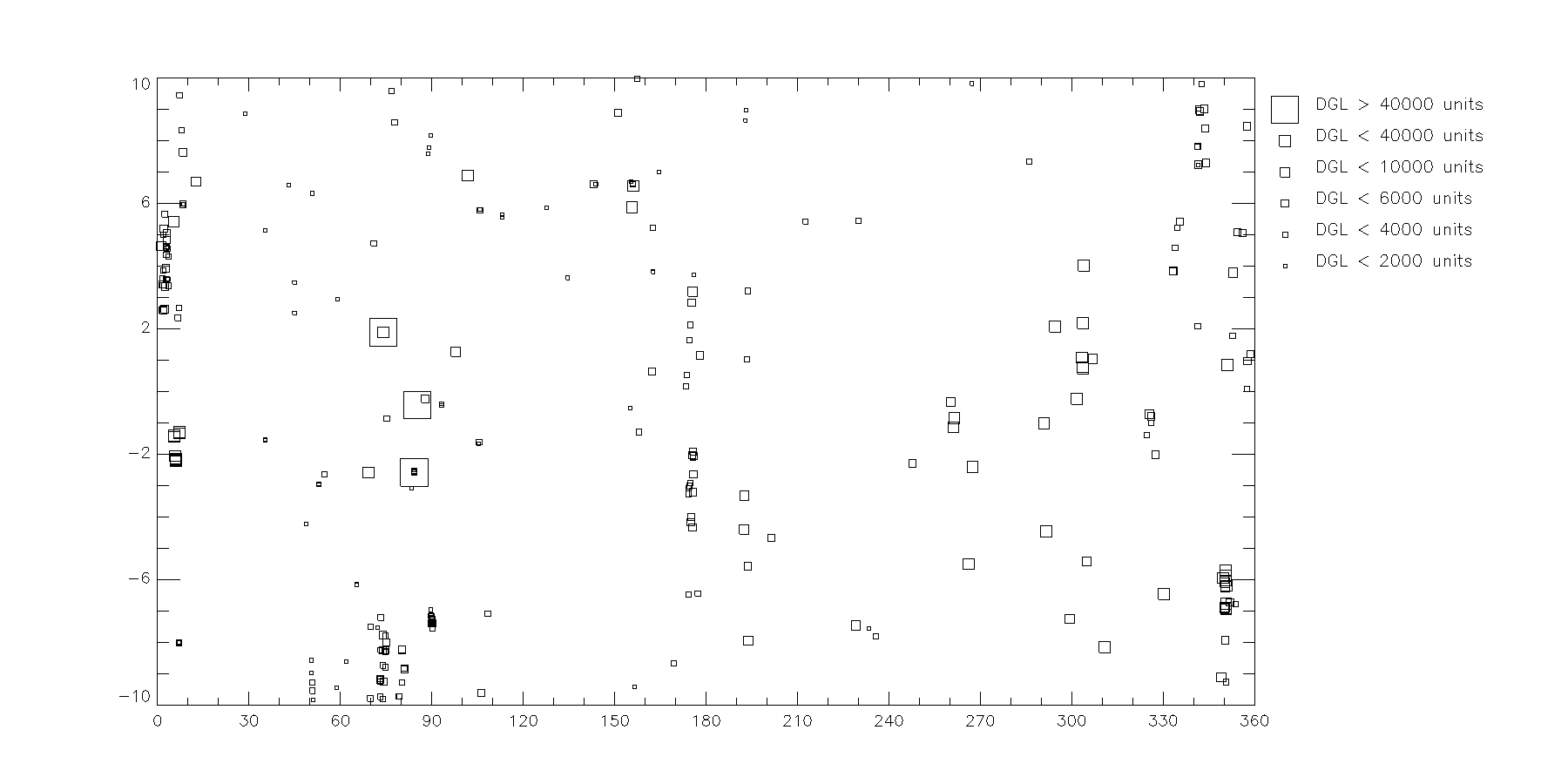}
    \caption{Locations of \voyager\ observations with the size of the symbol dependent on the observed surface brightness in \photu.}
    \label{fig:voy_pos}
\end{figure*}

\begin{figure}
	\includegraphics[width=\columnwidth]{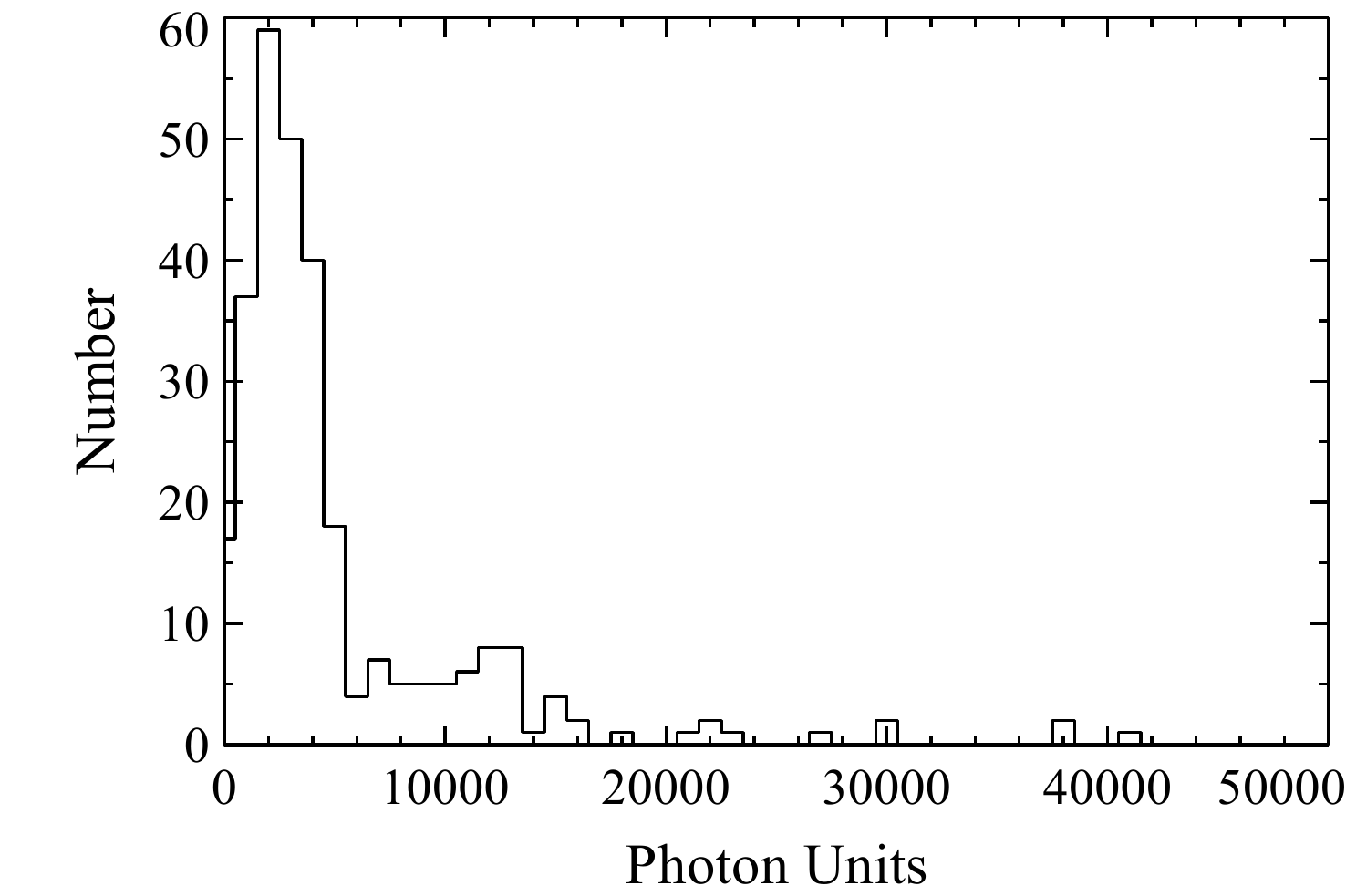}
    \caption{Histogram of observed surface brightness in \photu.}
    \label{fig:voy_hist}
\end{figure}

There were 292 background observations within $10^{\circ}$ of the Galactic plane  and we have plotted these in Fig. \ref{fig:voy_pos} where the size of the symbol is dependent on the mean surface brightness between 1070 and 1130 \AA. The distribution of the observed brightness is plotted in Fig. \ref{fig:voy_hist}. We emphasize that, even though these are observations near the Galactic Plane, most (206) of them have a surface brightness of less than 5000 \photu\ and only 47 with a surface brightness of more than 10,000 \photu. 

\begin{table}
	\centering
	\caption{Bright \voyager\ targets.}
	\label{tab:very_bright_voy}
	\begin{tabular}{llllll}
	\hline
	Spacecraft & Seq.$^{a}$ & Date & GL & GB & S$^{b}$\\ 
		\hline
		1 & 227 & 1990.68 & 85.22 & -0.43 & 117,000\\
		1 & 281 & 1994.07 & 84.28 & -2.59 & 51,800\\
		2 & 361 & 1985.76 & 74.05 & 1.89 & 41,500\\
		2 & 656 & 1991.32 & 156.12 & 6.57 & 38,761\\
		2 & 743 & 1992.83 & 294.55 & 2.08 & 30,121\\
		2 & 795 & 1993.44 & 155.69 & 5.87 & 15,394\\
		\hline
		\multicolumn{6}{l}{$^{a}$Sequence number from \citet{Murthyvoy_all}.}\\
		\multicolumn{6}{l}{$^{b}$Surface brightness in \photu\ from \citet{Murthyvoy_all}.}\\
	\end{tabular}
\end{table}

\begin{figure}
	\includegraphics[width=\columnwidth]{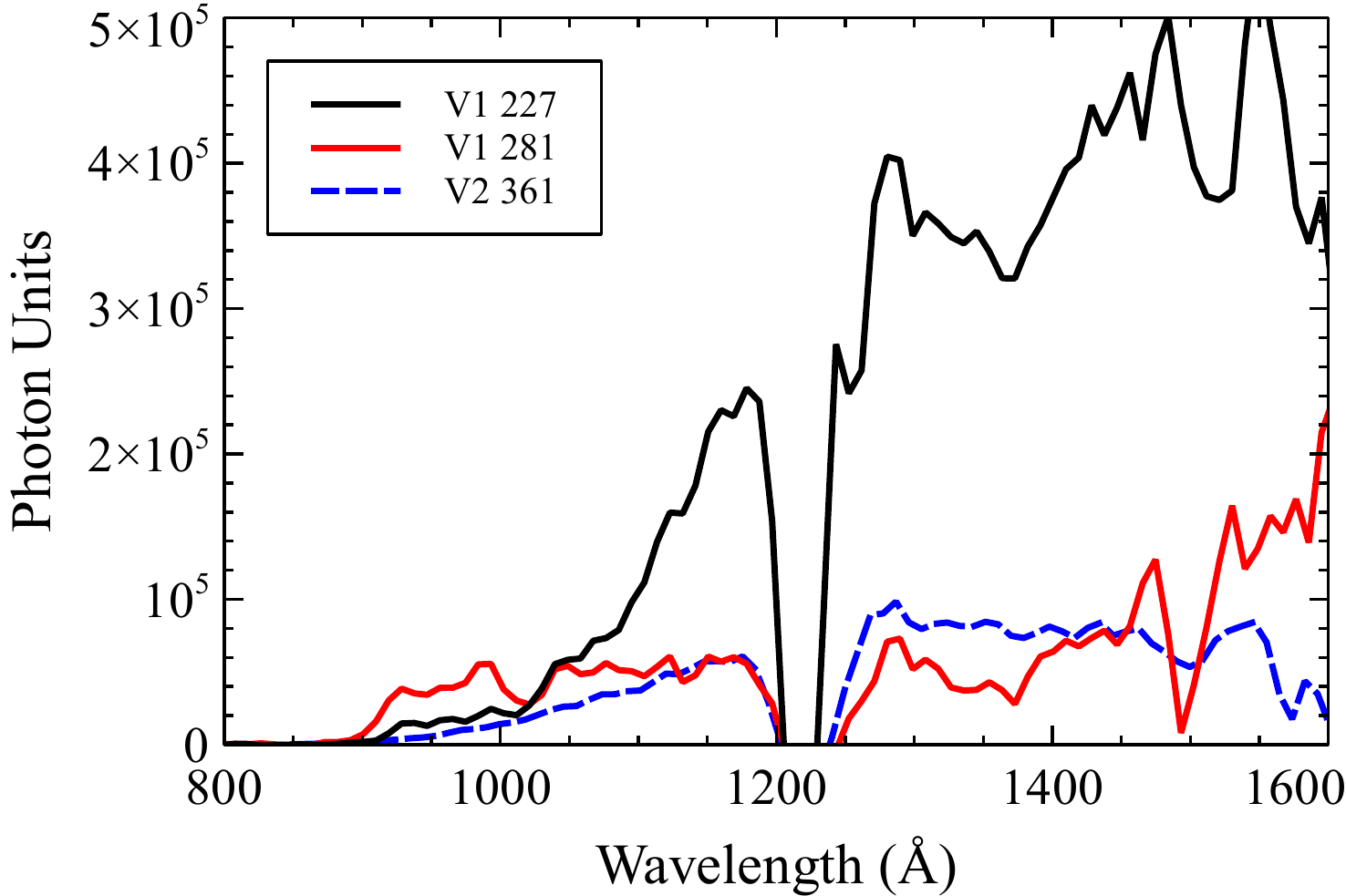}
    \caption{Three brightest \voyager\ observations in the Galactic plane.}
    \label{fig:voy_bright}
\end{figure}

There were several bright observations in the \voyager\ targets in the Galactic plane (Table \ref{tab:very_bright_voy}). All are consistent with an interstellar diffuse source (see \citet{Murthyvoy_all} for details) but exist in isolation; that is, there is no plausible source in optical maps of the region. A few even overlap with the \galex\ data but, again, with no evidence for such a bright source. Given the passage of time since the \voyager\ observations, we have not included them in this work and have left out a few others where the background subtraction was unreliable. This left 282 of the original 292 observations. The spectra for all \voyager\ diffuse observations, including those left out of this work, may be obtained from \citet{Murthyvoy_all}.

\subsection{\galex}

\begin{figure*}
	\includegraphics[width=6in, height=1.33in]{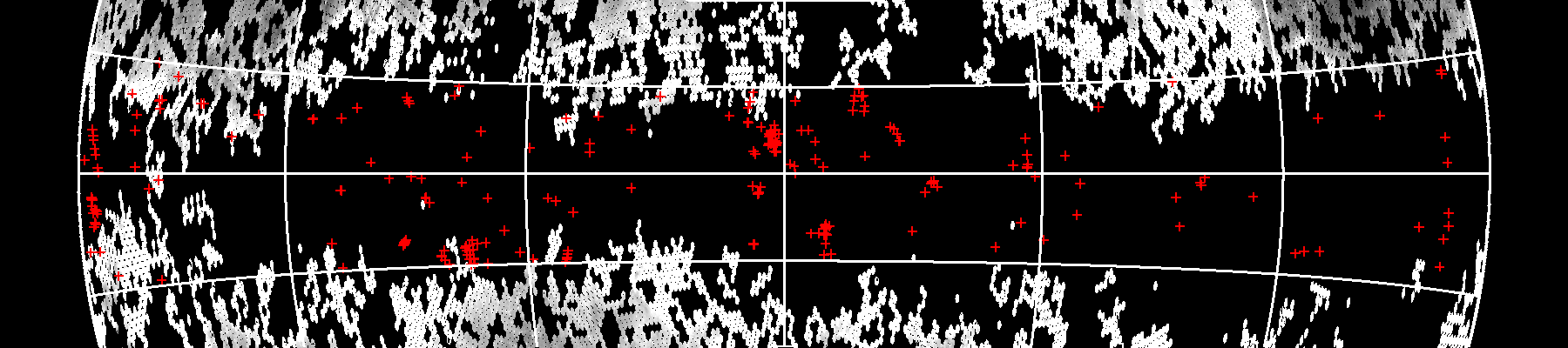}
	\includegraphics[width=6in, height=1.33in]{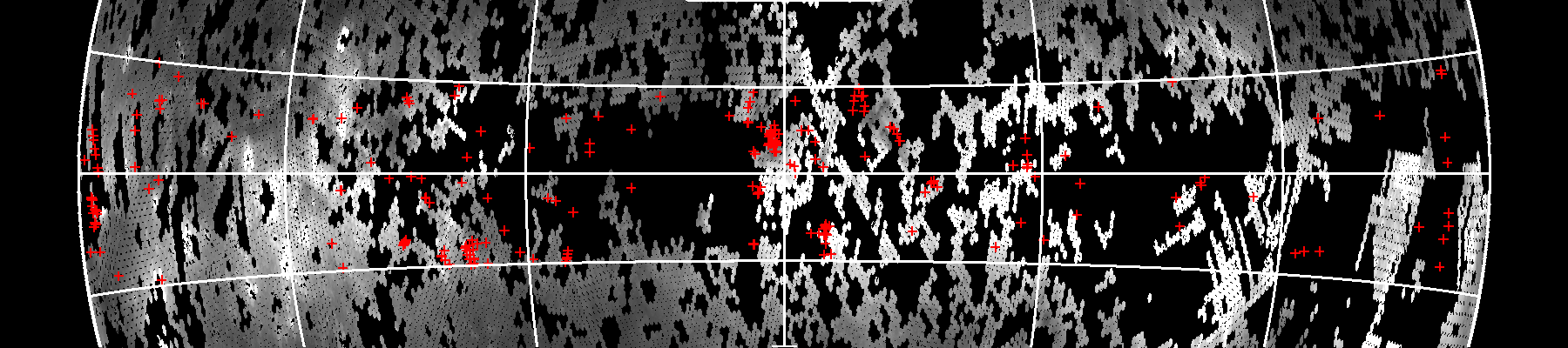}
    \caption{FUV (top) and NUV (bottom) images of the Galactic plane from \galex. The scale in both is linear with a maximum of 2000 \photu\ in the FUV and 5000 \photu\ in the NUV. The Galactic origin is at the center of the plot with longitude intervals every $60^{\circ}$ and latitude intervals at $10^{\circ}$. \voyager\ observations are shown as plus signs}
    \label{fig:galex_sky}
\end{figure*}

The \galex\ mission was launched in 2003 and continued to take observations until 2013 \citep{Martin2005,Morrissey2007}. There were two imagers on board: the far ultraviolet (FUV: 1516 \AA), which failed in 2007, and the near ultraviolet (NUV: 2316 \AA). Observations of the Galactic plane were barred for much of the mission under the mistaken belief that the diffuse radiation was intense, as would have predicted from a cosecant law \citep{Murthy_galex_data2010}, and only started near the end of the mission. Each \galex\ observation is an image of a $1.2^{\circ}$ field in the sky with a spatial resolution of 5\arcsec. The typical observation length was 100 seconds but could be much longer in a few fields. \citet{Murthy2014apss} used the original \galex\ data, removed the stars, subtracted the foreground airglow and zodiacal light \citep{Murthy2014apss}, and binned the data to produce a map of the diffuse radiation over the entire sky with a bin size of 6\arcmin. We have extracted the FUV and NUV observations within $10^{\circ}$ of the plane from these data (Fig. \ref{fig:galex_sky}. 

\begin{figure}
	\includegraphics[width=\columnwidth]{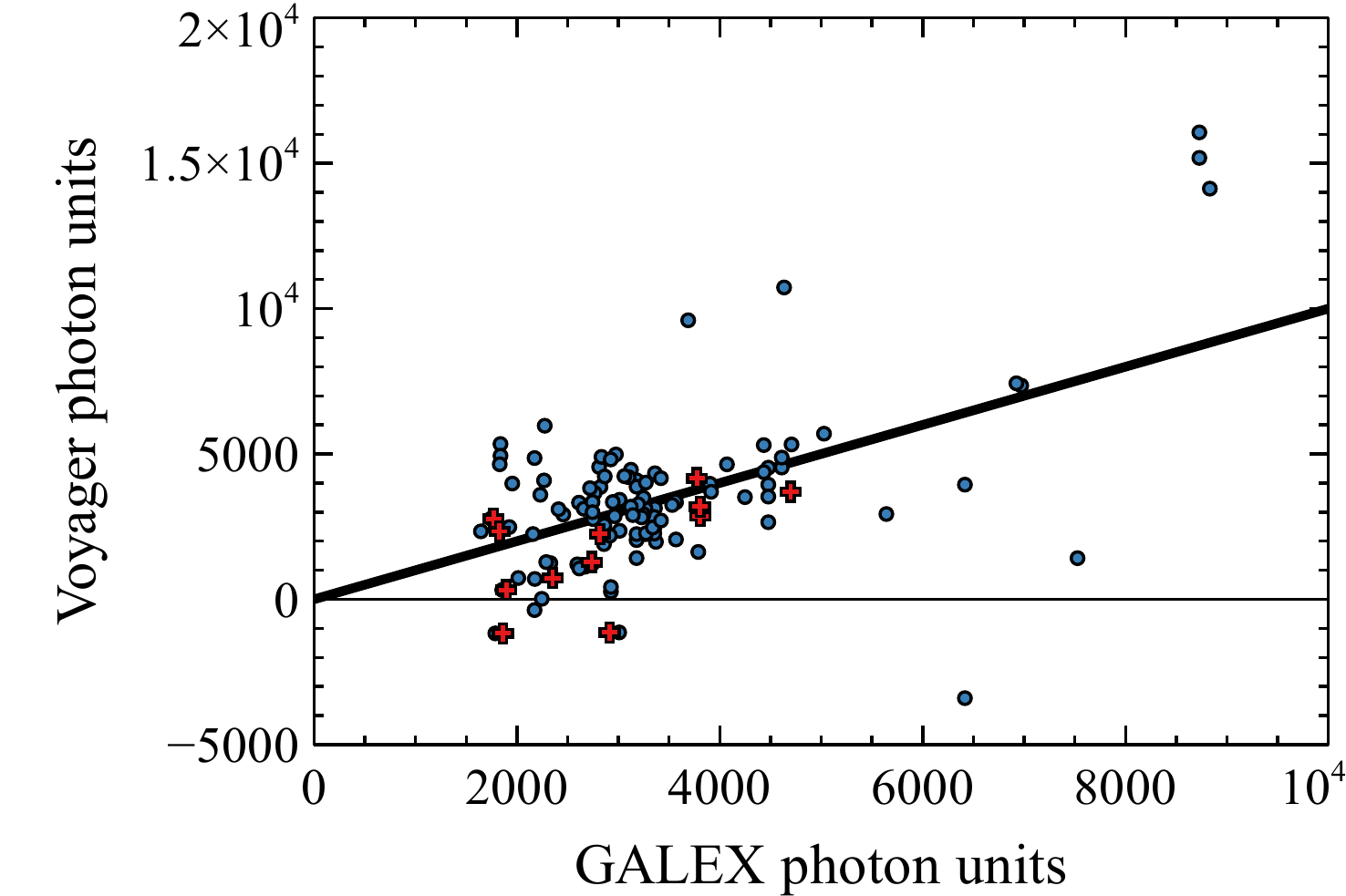}
    \caption{\voyager\ UVS observations at 1100 \AA\ versus \galex\ FUV (red circles) and NUV (plus signs) surface brightness. The two lines represent x = y (dark line) and y = 0.}
    \label{fig:voy_galex}
\end{figure}

As discussed above, the Galactic plane was only observed near the end of the \galex\ mission and only 125 of the \voyager\ targets are covered by NUV observations with 19 covered by FUV observations. There is a reasonable correlation between the \voyager\ DGL observations at 1100 \AA\ and the \galex\ FUV and NUV observations (Fig. \ref{fig:voy_galex}), with the outliers already discussed (Table \ref{tab:very_bright_voy}).

\section{Modeling \& Analysis}

\begin{figure}
	\includegraphics[width=\columnwidth]{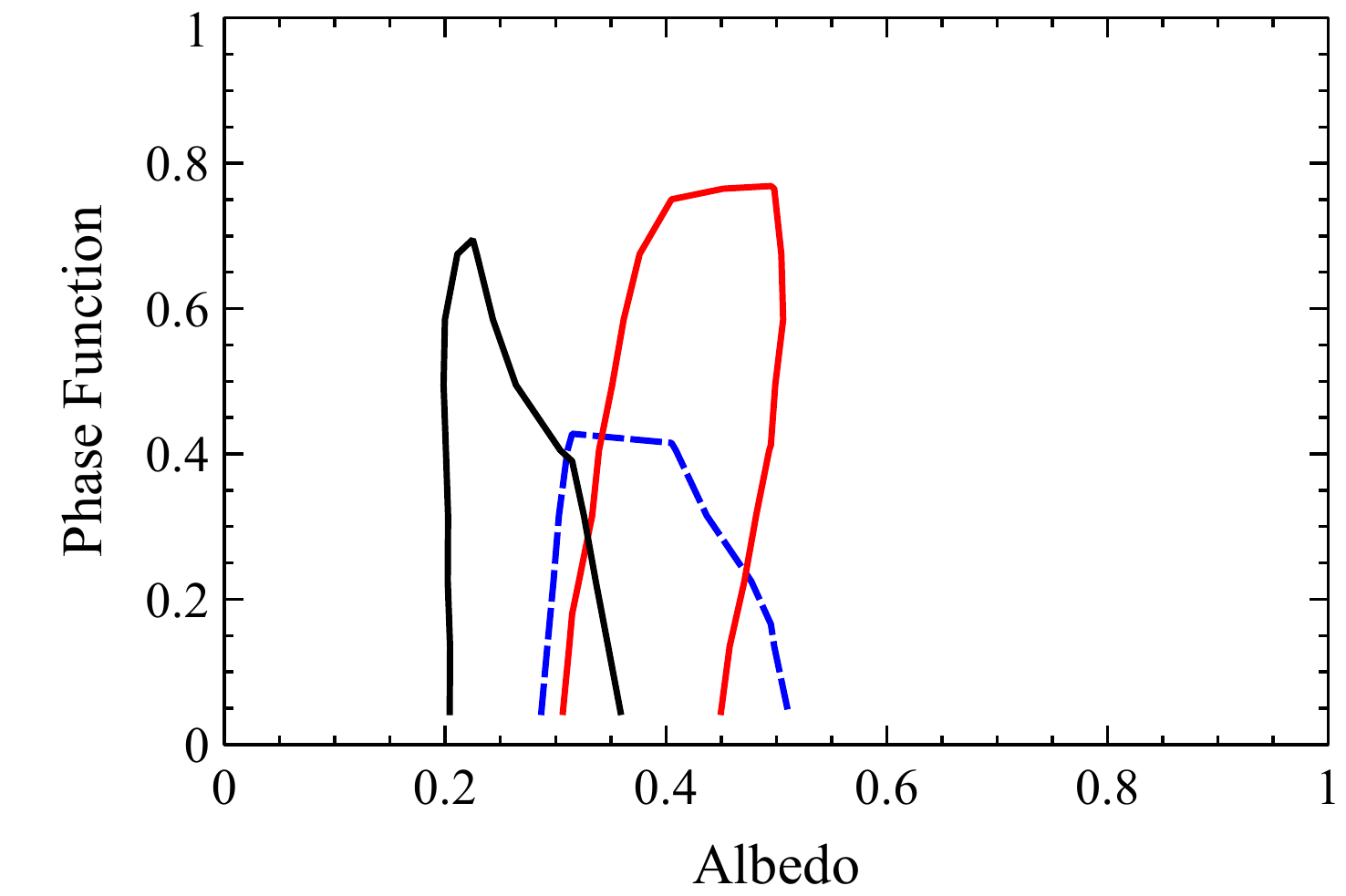}
    \caption{Allowed regions for optical constants of dust at 1100 \AA\ (black solid line), 1500 \AA\ (red solid line) and 2300 \AA\ (blue dashed line). }
    \label{fig:phase_albedo}
\end{figure}

\citet{Murthy2015, Murthy_dustmodel2016} developed a Monte Carlo model in which they used:
\begin{itemize}
    \item The stellar positions, spectral types, magnitudes and positions from the Hipparcos catalog \citep{Perryman1997}.
    \item Stellar spectra from \citet{Castelli2004}.
    \item A 3-dimensional dust model using dust maps from \citet{Schlegel1998} with an exponential drop off from the Galactic plane.
    \item A scattering phase function from \citet{Henyey1941} with the albedo ($a$) and phase function asymmetry factor ($g = <cos\theta>$) as free parameters. 
\end{itemize}
This model tracked the observed emission well at high Galactic latitudes but did increasingly poorly at lower latitudes as the dust column density increased. The dust cross-section  at 100 \micron\ is much less than that in the UV and the 100 \micron\ emission from the {\it Infrared Astronomy Satellite} ({\it IRAS}: \citet{neugebauer_iras1984}) samples a much longer line of sight than would be the case in the UV where the DGL comes primarily from dust within a few hundred parsecs of the Sun \citep{Murthy_dustmodel2016}. We have therefore replaced the 2-dimensional dust map from \citet{Schlegel1998} with the 3-d map of \citet{Green2015}, as described by \citet{Akshaya2019}.

\begin{table}
	\centering
	\caption{Optical constants.}
	\label{tab:albedo_phase}
	\begin{tabular}{lll}
	\hline
	Wavelength (\AA ) & Albedo ($a$) & Phase Function ($g$)\\ 
		\hline
		1100 & 0.2 -- 0.3 & $ < 0.7$\\
		1500 & 0.2 -- 0.5 & $ < 0.8$\\
		2300 & 0.4 -- 0.6 & $ < 0.4$\\
		\hline
	\end{tabular}
\end{table}

We have fit our observations with our models at 1100, 1500, and 2300 \AA\ as a function of the optical constants ($a$ and $g$) and have drawn pseudo-confidence contours in Fig. \ref{fig:phase_albedo} to find limits on the optical constants (Table \ref{tab:albedo_phase}). Our models are time consuming and we have only run them over a limited grid of optical parameters with a spacing of 0.1 on both $a$ and $g$. Nevertheless, we find limits that are consistent with others \citep{Draine2003, Akshaya2018}. 

\begin{figure}
	\includegraphics[width=\columnwidth]{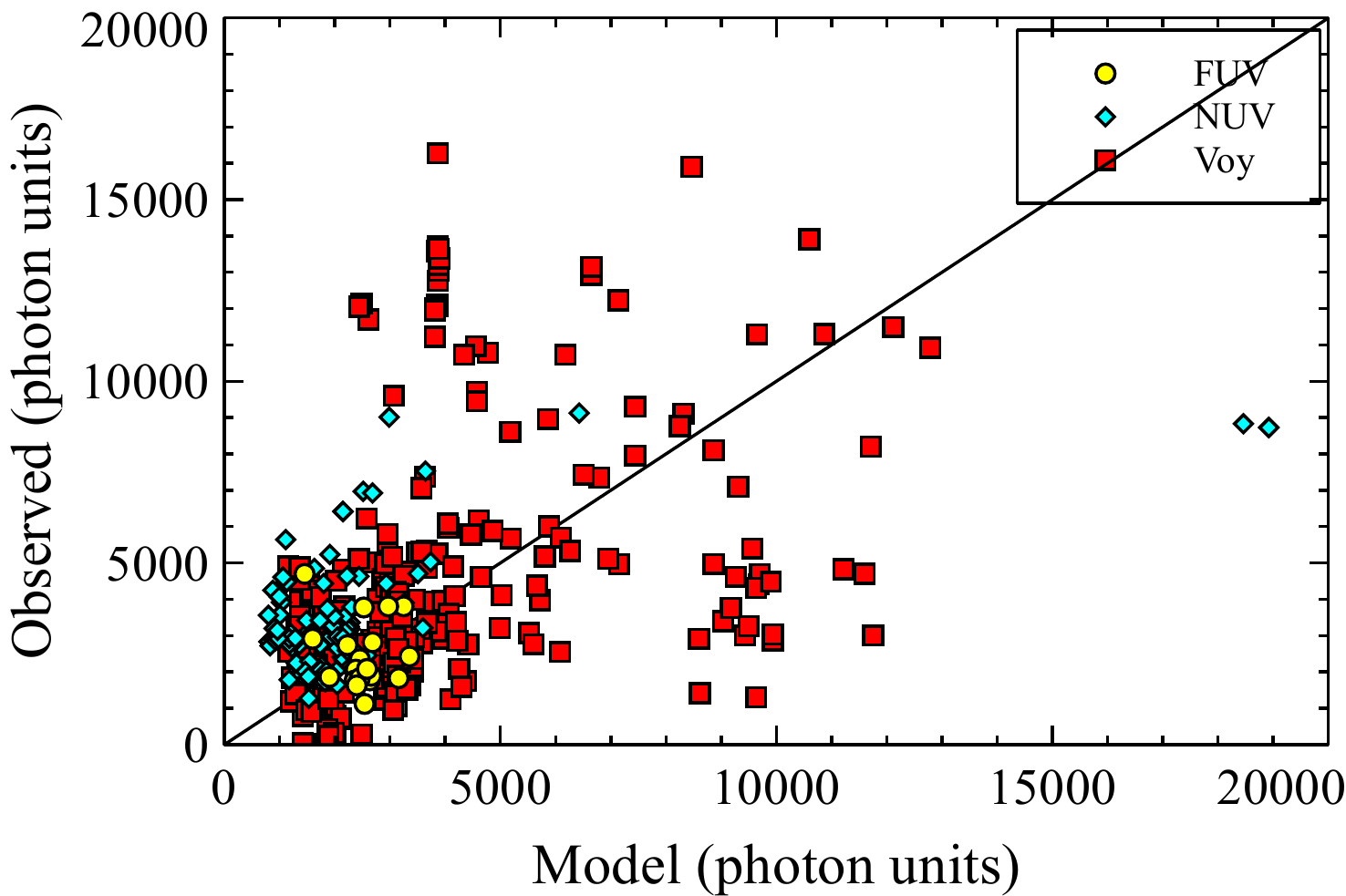}
    \caption{Data from \voyager\ (1100 \AA) \galex\ FUV (1500 \AA) and NUV (2300 \AA) versus model predictions. The line has a slope of 1 and is shown to guide the eye.}
    \label{fig:data_model}
\end{figure}

The observations have been plotted against the models ($a = 0.4$; $g = 0.4$) for each of the three bands in Fig. \ref{fig:data_model}. We find that the models are a reasonable fit to the data in most cases with some notable deviations. There are several reasons why this could arise. Perhaps the most critical is that the DGL in the UV is dependent on local effects \citep{Murthy_sahnow2004} and we will have to model each direction individually. We also plan to explore the parameter space with a better appreciation of the errors in both the modeling and the data and a finer grid in the optical parameters.

\section{Conclusions}

We have used archival observations from \voyager\ and \galex\ to study the DGL near the Galactic plane. Although there are a few particularly bright regions for which we have no explanation, most of the observations are a few thousand \photu\ from 1100 \AA\ to 2300 \AA. We will examine the bright regions further in a future work but expect that it will be difficult to find a plausible explanation without further observations.

We have compared a model for dust scattering in the Galaxy \citep{Murthy_dustmodel2016} to the observations finding a reasonable fit but with several outliers. These are perhaps because of local effects which we will explore in a future work. We find an albedo of 0.2 -- 0.3 and $g < 0.7$ at 1100 \AA; 0.2 -- 0.5 and $g < 0.8$ at !500 \AA; and 0.4 -- 0.6 at 2300 \AA.
We plan now to examine individual regions in more detail to understand the distribution of dust in those directions and, perhaps, changes in the optical constants, if any.

\section*{Acknowledgements}

We have used the Gnu Data Language (http://gnudatalanguage.sourceforge.net/index.php) for the analysis of these data. The data presented in this paper were obtained from the Mikulski Archive for Space Telescopes (MAST). STScI is operated by the Association of Universities for research in Astronomy, Inc., under the NASA contract NAS5-26555.  Support for MAST for non-HST data is provided by the NASA Office of Space
Science via grant NNX09AF08G and by other grants and contracts. This work has been done while under work-at-home rules and we thank all those who have made it possible for us to do so.




\bibliographystyle{mnras}



\appendix

\section{Some extra material}

If you want to present additional material which would interrupt the flow of the main paper,
it can be placed in an Appendix which appears after the list of references.


\bsp	
\label{lastpage}
\end{document}